\documentclass[12pt]{article}
\usepackage[margin=0.7in]{geometry}

\usepackage{amsmath}
\usepackage{amsfonts}
\usepackage{amsthm}
\usepackage{array}
\usepackage{multirow}
\usepackage{graphicx}
\usepackage{caption}
\usepackage{subcaption}

\DeclareMathOperator{\grad}{grad}

\newcommand{\totalDiff}[1]{\frac{\mathrm{d}}{\mathrm{d}{#1}}}
\newcommand{\parder}[2]{\frac{\partial #1}{\partial #2}}

\newcommand{\dir}[1]{\partial_{ {#1}}}

\newcommand{\LieDerivative}[2]{\mathcal{L}_{#1}\left(#2\right)}

\newcommand{\LieAlgebra}[1]{\mathfrak{#1}}

\newcommand{\bvec}[1]{\mathbf{#1}}
\newcommand{\entr}{s}
\newcommand{\temp}{T}
\newcommand{\dens}{\rho}
\newcommand{\press}{p}
\newcommand{\energy}{\epsilon}

\newcommand{\systemEk}[1]{\mathcal{E}_{#1}}

\newtheorem{remark}{Remark}
\newtheorem{theorem}{Theorem}

\title{Symmetries and Differential Invariants for Inviscid Flows on a Curve}
\author{Anna Duyunova,
	\\ Institute of Control Sciences of RAS,
	\\ duyunova\_anna@mail.ru,\\
	Valentin Lychagin,
	\\ Institute of Control Sciences of RAS, University of Troms\o,\\valentin.lychagin@uit.no,\\
	Sergey Tychkov,
	\\ Institute of Control Sciences of RAS,\\ sergey.lab06@yandex.ru
}
\date{}

\begin{document}
\maketitle 	
\abstract{ Symmetries and the corresponding fields
of differential invariants 
of the inviscid flows on a curve are given.
Their dependence on thermodynamic states
of media is studied, and a classification of thermodynamic
states is given.}

\section{Introduction}


Consider flows of an inviscid medium on an oriented Riemannian manifold $M$ with a structure form $g$ in the field of constant gravitational force. Such flows satisfy the Euler system consisting of the following equations (see \cite{Batchelor2000} for details):

\begin{equation}\label{eq:E}
\left\{
\begin{aligned}
& \dens(\bvec{u}_t  + \nabla_{\bvec{u}}\bvec{u})=- \grad{\press} +  \bvec{g}\dens ,\\
& \parder{(\dens\, \Omega_g)}{t} + \LieDerivative{\bvec{u}}{\dens\, \Omega_g} = 0,\\
&\temp\left(  \entr_t + \nabla_{\bvec{u}}s\right)  - \frac{k}{\dens} \Delta_g T =0,
\end{aligned}
\right.
\end{equation}
where the vector field $\bvec{u}=(u,v,w)$
is the flow velocity, $\press$, $\dens$,
$\entr$, $\temp$ are the pressure, density,
specific entropy, temperature of the fluid  
respectively, $k$ is the thermal conductivity, which is supposed to be constant, and $\bvec{g}$ is the gravitational acceleration.

Here $\nabla_X$ is the directional covariant Levi-Civita derivative with respect to a vector field $X$, $\Omega_g$ is the volume form on the manifold $M$, $\Delta_{g}$ is the Laplace-Beltrami operator corresponding to the metric $g$.

We consider a flow on a naturally-parameterized curve 
\[
M = \{x=f(a),\,  y=g(a), \, z=h(a)\} 
\]
in the three-dimensional Euclidean space. In this case vector  $\bvec{g}$ is the restriction of the vector field $(0,0,\mathrm{g})$ on $M$, i.e. 
\[
\bvec{g} =- \mathrm{g}  h^{\prime} \partial_a.
\]

First of all, we note that the system~\eqref{eq:E} is incomplete, namely, it lacks two additional relations between thermodynamic quantities. To obtain them, we employ the same method as we did in the paper \cite{DLTwisla}. 
The idea of this method is based on interpretation of media thermodynamic states as Legendrian, or Lagrangian, manifolds in contact, or symplectic, space correspondingly.

So, by the system  $\systemEk{}$ of differential equations, we mean the differential equations \eqref{eq:E} and two equations of the thermodynamic state 
\begin{equation}\label{eq:Therm}
L= \{\, F(\press,\dens,\entr,\temp)=0, \, G(\press,\dens,\entr,\temp)=0 \,\}
\end{equation}
that satisfy the relation 
\[
[F,G]=0 \quad \mathrm{mod} \quad \{F=0,\,G=0\} ,
\]
where $[F,G]$ is the Poisson bracket with respect to the symplectic form 
\[
\Omega = d\entr \wedge d\temp + {\dens^{-2}} d\dens \wedge d\press .
\]
Moreover, the restriction of quadratic differential form
\[
\kappa= d(\temp^{-1})\cdot d\energy - \dens^{-2} d(\press \temp^{-1}) \cdot d\dens
\]
on the manifold of thermodynamic state is negative definite, here $\energy$ is the specific internal energy. 

The paper is organized as follows.

In Section \ref{sec:symmetries} we study symmetry Lie algebras of the Euler system $\systemEk{}$  and their dependence on the form of the function $h(a)$. There are six different forms, besides the general case, of the function $h$ that correspond to different symmetry algebras.

In Section \ref{sec:lift} we discuss representation of a space curve
as a lift of a plane curve, as well as connection between the function $h$ and a way of lifting curve. For each case of $h$ described in Section \ref{sec:symmetries}, we present form of the `lifting' function explicitly.  As examples, we demonstrate lift of the unit circle. 

In Section \ref{sec:therm} we consider the thermodynamic states and the corresponding
Lie algebras for the cases when the thermodynamic state
admits a one-dimensional symmetry algebra.
For such thermodynamic states,
we find an explicit form of Lagrangian surface
in terms of two relations between the thermodynamic
quantities $\press$, $\temp$, $\dens$ and $\entr$.

In Section \ref{sec:invs} we recall briefly 
the notion of differential invariants and introduce two types of invariants considered in this paper, namely, Euler and kinematic invariants. These types differ in a Lie algebra with respect to the action of which they are invariant. For both types
we provide a full description of the field of invariants including
basis invariants and invariant derivations. Additionally, for the Euler invariants, special cases of the function $h$ are considered.

Most of the computations in this paper were done in Maple with
the Differential Geometry package by I. Anderson and his team \cite{Anderson2016}.


\section{Symmetry Lie algebra}\label{sec:symmetries}

Using the standard techniques of symmetries computation we obtain  
that, depending on the function $h(a)$, the symmetry algebra of system $\systemEk{}$ has different generators (see the Maple files \textit{http://d-omega.org}).

To describe this Lie algebra,
we consider a Lie algebra $\LieAlgebra{g}$ of point symmetries of the PDE system \eqref{eq:E}.


Let $\vartheta :\LieAlgebra{g} \rightarrow \LieAlgebra{h}$ be
the following Lie algebras homomorphism
\[
\vartheta : X\mapsto
X(\dens)\dir{\dens} + X(\entr)\dir{\entr} + X(\press)\dir{\press} + X(\temp)\dir{\temp},
\]
where $\LieAlgebra{h}$ is a Lie algebra generated by vector fields that act on the thermodynamic valuables $\press$, $\dens$, $\entr$ and $\temp$.

The kernel of the homomorphism $\vartheta$ is an ideal  $\LieAlgebra{g_{m}}\subset \LieAlgebra{g}$, and we call the elements of $\LieAlgebra{g_{m}}$ \textit{geometric symmetries}.

Let also $\LieAlgebra{h_{t}}$ be the Lie subalgebra of the algebra $\LieAlgebra{h}$
that preserves thermodynamic state \eqref{eq:Therm}.

Then the following result is valid (see for details \cite{DLTwisla}).
\begin{theorem}
	A Lie algebra $\LieAlgebra{g_{sym}}$ of symmetries of the Euler system $\systemEk{}$ coincides with 
	\[
	\vartheta^{-1}(\LieAlgebra{h_{t}}).
	\]
\end{theorem}

First of all, consider the general case, when $h(a)$ is an arbitrary function. Then the Lie algebra $\LieAlgebra{g^0}$ of point symmetries of the system \eqref{eq:E} is generated by the vector fields 
\begin{equation*} 
	\begin{aligned} 
   &X_1 = \dir{ t}, \qquad  
   X_{4}=\temp\,\dir{ \temp}, \\
   &X_{2} = \dir{ \press} , \qquad  
   X_{5} = \press\,\dir{ \press} + \dens\,\dir{ \dens} -\entr\,\dir{ \entr}  .\\
   &X_{3} =  \dir{ \entr}  ,
	\end{aligned} 
\end{equation*}

Note that the Lie algebra $\LieAlgebra{g^0}$ is solvable and the sequence of derived algebras is the following
\[
\LieAlgebra{g^0} = \left\langle X_1,X_2,\ldots, X_5 \right\rangle
\supset
\left\langle X_2,X_3 \right\rangle=0. 
\]

The pure thermodynamic part $\LieAlgebra{h^0}$ of the system symmetry algebra in this case is generated by
\begin{equation*} 
\begin{aligned} 
&Y_1 = \dir{ \press}, \qquad  
Y_{3}=\temp\,\dir{ \temp}, \\
&Y_{2} = \dir{ \entr} , \qquad  
Y_{4} = \press\,\dir{ \press} + \dens\,\dir{ \dens} -\entr\,\dir{ \entr}  .
\end{aligned} 
\end{equation*}

Thus, the system of differential equations $\systemEk{}$ has the smallest Lie algebra of point symmetries $ \vartheta^{-1}(\LieAlgebra{h_{t}^0})$, when the function $h(a)$ is arbitrary.

Below we list special cases of the function $h(a)$.

\medskip
\textbf{1.} $h(a)= const$ 

In this case the Lie algebra $\LieAlgebra{g^1}$ of point symmetries of the system \eqref{eq:E} is generated by the vector fields $X_1,X_2,\ldots,X_5$ and by the following vector fields
\[
\begin{aligned} 
&X_6 = \dir{ a}, \qquad \qquad \,\,\,\,
X_8 = t\,\dir{ t}+a\,\dir{a}-\entr\,\dir{\entr}, \\
&X_7 = t\,\dir{ a}+\dir{ u}, \qquad
X_{9} = t\,\dir{ t} - u\,\dir{ u} -2 \press\,\dir{ \press} + \entr\,\dir{ \entr}.
\end{aligned} 
\]

The Lie algebra $\LieAlgebra{g^1}$ is solvable and the sequence of derived algebras is the following 
\[
 \LieAlgebra{g^1} = \left\langle X_1,X_2,\ldots,X_9 \right\rangle \supset
\left\langle X_1,X_2,X_3,X_6,X_7 \right\rangle \supset
\left\langle X_6\right\rangle  =0.
\]

The pure thermodynamic part $\LieAlgebra{h^1}$ of the symmetry algebra is generated by the vector fields 
\begin{equation*} 
\begin{aligned} 
&Y_1 = \dir{ \press}, \qquad  
Y_{3}=\temp\,\dir{ \temp}, \qquad
Y_{5} =  \dens\,\dir{ \dens}, \\
&Y_{2} = \dir{ \entr} , \qquad  
Y_{4} = \press\,\dir{ \press} ,\qquad \,\,
Y_{6} = \entr\,\dir{ \entr}  .
\end{aligned} 
\end{equation*}

Thus, the system of differential equations $\systemEk{}$ has a Lie algebra of point symmetries 
$
 \vartheta^{-1}(\LieAlgebra{h_{t}^1}).
$

\medskip
\textbf{2.} $h(a)= \lambda a$,  $\lambda\neq 0$

In this case the Lie algebra $\LieAlgebra{g^2}$ of point symmetries of the system \eqref{eq:E} is generated by the vector fields $X_1,X_2,\ldots,X_5$ and by the following vector fields
\[
\begin{aligned} 
&X_6 = \dir{ a}, \qquad\\
&X_7 = t\,\dir{ a}+\dir{ u}, \qquad \\
&X_8 = t\,\dir{ t}+2a\,\dir{a} +  u\,\dir{ u} -2 \dens\,\dir{ \dens} -\entr\,\dir{\entr}, \\
&X_{9} = \left(  \frac{t^2}{2}+\frac{a}{\lambda \mathrm{g}}\right) \dir{ a} +\left( t+\frac{u}{\lambda \mathrm{g}} \right) \dir{ u} - 
\frac{2 \dens }{\lambda \mathrm{g}}\,\dir{ \dens} .
\end{aligned} 
\]

The Lie algebra $\LieAlgebra{g^2}$ is solvable and the sequence of derived algebras is the following 
\[
\LieAlgebra{g^2} = \left\langle X_1,X_2,\ldots,X_9 \right\rangle \supset
\left\langle X_1,X_2,X_3,X_6,X_7 \right\rangle \supset
\left\langle X_6\right\rangle =0.
\]

The pure thermodynamic part $\LieAlgebra{h^2}$ of the symmetry algebra is generated by the vector fields 
\begin{equation*} 
\begin{aligned} 
&Y_1 = \dir{ \press}, \qquad  
Y_{3}=\temp\,\dir{ \temp}, \qquad
Y_{5} =  \dens\,\dir{ \dens}, \\
&Y_{2} = \dir{ \entr} , \qquad  
Y_{4} = \press\,\dir{ \press} ,\qquad \,\,
Y_{6} = \entr\,\dir{ \entr}  .
\end{aligned} 
\end{equation*}

Thus, the system of differential equations $\systemEk{}$ has a Lie algebra of point symmetries
$
 \vartheta^{-1}(\LieAlgebra{h_{t}^2}).
$

\medskip
\textbf{3.} $h(a)= \lambda a^2$, $\lambda\neq 0$

In this case the Lie algebra $\LieAlgebra{g^3}$ of point symmetries of the system \eqref{eq:E} is generated by the vector fields $X_1,X_2,\ldots,X_5$ and, if $\lambda>0$, by the vector fields 
\[
\begin{aligned} 
&X_6 = a\, \dir{ a} +u\,\dir{ u}  -2 \dens\,\dir{ \dens} , \qquad\\
&X_7 = \sin(\sqrt{2\lambda \mathrm{g}}\,t)\,\dir{a} +  \sqrt{2\lambda \mathrm{g}} \cos(\sqrt{2\lambda \mathrm{g}}\,t)\,\dir{ u} , \qquad \\
&X_8 =\cos(\sqrt{2\lambda \mathrm{g}}\,t)\,\dir{a} -  \sqrt{2\lambda \mathrm{g}} \sin(\sqrt{2\lambda \mathrm{g}}\,t)\,\dir{ u} , \\
\end{aligned} 
\]
and, if $\lambda<0$,  by the vector fields 
\[
\begin{aligned} 
&X_6 = a\, \dir{ a} +u\,\dir{ u}  -2 \dens\,\dir{ \dens} , \qquad\\
&X_7 = \exp(\sqrt{-2\lambda \mathrm{g}}\,t)\,\dir{a} +  \sqrt{-2\lambda \mathrm{g}} \exp(\sqrt{-2\lambda \mathrm{g}}\,t)\,\dir{ u} , \qquad \\
&X_8 =\exp(-\sqrt{-2\lambda \mathrm{g}}\,t)\,\dir{a} -  \sqrt{-2\lambda \mathrm{g}} \exp(-\sqrt{-2\lambda \mathrm{g}}\,t)\,\dir{ u} . \\
\end{aligned}  
\]

The Lie algebra $\LieAlgebra{g^3}$ is solvable and the sequence of derived algebras is the following 
\[
\LieAlgebra{g^3} = \left\langle X_1,X_2,\ldots,X_8 \right\rangle\supset
\left\langle X_2,X_3,X_7,X_8 \right\rangle =0.
\]

The pure thermodynamic part $\LieAlgebra{h^3}$ of the symmetry algebra is generated by the vector fields 
\begin{equation*} 
\begin{aligned} 
&Y_1 = \dir{ \press}, \qquad  
Y_{3}=\temp\,\dir{ \temp}, \qquad 
Y_{5} =\press\,\dir{ \press} - \entr\,\dir{ \entr} .\\
&Y_{2} = \dir{ \entr} , \qquad  
Y_{4} = \dens\,\dir{ \dens},    
\end{aligned} 
\end{equation*}

Thus, the system of differential equations $\systemEk{}$ has a Lie algebra of point symmetries
$
 \vartheta^{-1}(\LieAlgebra{h_{t}^3}).
$

\medskip
\textbf{4.} $h(a)= \lambda_1a^{\lambda_2}$, $\lambda_2\neq 0,1,2$ 

The Lie algebra $\LieAlgebra{g^4}$ of point symmetries of the system \eqref{eq:E} is generated by the vector fields $X_1,X_2,\ldots,X_5$ and by the vector field
\[
X_6 = t\,\dir{ t} - \frac{2 a}{\lambda_2-2} \, \dir{ a} - \frac{\lambda_2 u}{\lambda_2-2}\,\dir{ u} 
+\frac{2\lambda_2 \dens}{\lambda_2-2}\,\dir{ \dens} -
 \entr\,\dir{ \entr}.
\]

The Lie algebra $\LieAlgebra{g^4}$ is solvable and the sequence of derived algebras is the following 
\[
\LieAlgebra{g^4} = \left\langle X_1,X_2,\ldots,X_6 \right\rangle\supset
\left\langle X_1,X_2,X_3 \right\rangle=0 .
\]

The pure thermodynamic part $\LieAlgebra{h^4}$ of the symmetry algebra is generated by the vector fields 
\begin{equation*} 
\begin{aligned} 
&Y_1 = \dir{ \press}, \qquad  
Y_{3}=\temp\,\dir{ \temp}, \qquad
Y_{5} =\press\,\dir{ \press} + \dens\,\dir{ \dens} - \entr\,\dir{ \entr} .  \\
&Y_{2} = \dir{ \entr} , \qquad  
Y_{4} =  2\lambda_2 \dens\,\dir{ \dens} -(\lambda_2-2) \entr\,\dir{ \entr}   ,
\end{aligned} 
\end{equation*}

Thus, the system of differential equations $\systemEk{}$ has a Lie algebra of point symmetries
$
 \vartheta^{-1}(\LieAlgebra{h_{t}^4}).
$

\medskip
\textbf{5.} $h(a)=\lambda_1e^{\lambda_2a}, \, \lambda_2\neq 0$

In this case the Lie algebra $\LieAlgebra{g^5}$ of point symmetries of the system \eqref{eq:E} is generated by the vector fields $X_1,X_2,\ldots,X_5$ and by the vector field
\[
X_6 = t\,\dir{ t}-\frac{2	}{\lambda_2}\,\dir{a} -  u\,\dir{ u} - \press\,\dir{ \press} +\dens\,\dir{\dens}.
\]

The Lie algebra $\LieAlgebra{g^5}$ is solvable and the sequence of derived algebras is the following 
\[
\LieAlgebra{g^5} = \left\langle X_1,X_2,\ldots,X_6 \right\rangle\supset
\left\langle X_1,X_2,X_3 \right\rangle=0 .
\]

The pure thermodynamic part $\LieAlgebra{h^5}$ of the symmetry algebra is generated by the vector fields 
\begin{equation*} 
\begin{aligned} 
&Y_1 = \dir{ \press}, \qquad  
Y_{3}=\temp\,\dir{ \temp}, \qquad
Y_{5} = 2 \dens\,\dir{ \dens} - \entr\,\dir{ \entr} .\\
&Y_{2} = \dir{ \entr} , \qquad  
Y_{4} =  \press\,\dir{ \press} -\dens\,\dir{\dens} ,
\end{aligned} 
\end{equation*}

The system of differential equations $\systemEk{}$ has a Lie algebra of point symmetries
$
 \vartheta^{-1}(\LieAlgebra{h_{t}^5}).
$

\medskip
\textbf{6.} $h(a)= \ln a$

The Lie algebra $\LieAlgebra{g^6}$ of point symmetries of the system \eqref{eq:E} is generated by the vector fields $X_1,X_2,\ldots,X_5$ and by the vector field
\[
X_6  = t\,\dir{ t}+ a  \,\dir{a}-\entr\,\dir{\entr}.
\]

The Lie algebra $\LieAlgebra{g^6}$ is solvable and the sequence of derived algebras is the following 
\[
\LieAlgebra{g^6} = \left\langle X_1,X_2,\ldots,X_6 \right\rangle \supset
\left\langle X_1,X_2,X_3 \right\rangle =0.
\]

The pure thermodynamic part $\LieAlgebra{h^6}$ of the symmetry algebra is generated by the vector fields 
\begin{equation*} 
\begin{aligned} 
&Y_1 = \dir{ \press}, \qquad  
Y_{3}=\temp\,\dir{ \temp}, \qquad
Y_{5} = \press\,\dir{ \press} + \dens\,\dir{ \dens} \\
&Y_{2} = \dir{ \entr} , \qquad  
Y_{4} =  \entr\,\dir{\entr }  ,  
\end{aligned} 
\end{equation*}

The system of differential equations $\systemEk{}$ has a Lie algebra of point symmetries
$
\vartheta^{-1}(\LieAlgebra{h_{t}^6}).
$

\begin{remark}
	Since we are allowed to choose any starting point on
	the curve and any horizontal plane to lift
	the curve from, symmetry algebras corresponding to the functions $h(a+a_0)$, $h(a)+h_0$ and $h(a)$ are the same.
\end{remark}

\section{Lifting curves from the plane} \label{sec:lift}

Consider geometrical interpretation of the each case for the function $h(a)$ given above.

Let a curve in the space be defined as a pair of a plane curve $(x(\tau), y(\tau))$ and a `lifting' function $z(\tau)$. Also, denote length of the plane curve $\int\limits_0^{\tau }\sqrt{x^2_{\theta}+y^2_{\theta}}\,d\theta$ by $l(\tau)$. Then the following relation
between natural parameter $a$ and the parameter $\tau$ is valid
\[
h_a = \frac{z_{\tau}}{ \sqrt{x^2_{\tau}+y^2_{\tau}+z^2_{\tau}}}   .
\]

We consider different ways of lifting a curve from the plane depending on the particular form of the function $h(a)$.


\medskip
\textbf{1.} $h(a)=const$

The first way of lifting a plane curve is to translate the whole curve along $z$-axis, i.e. if $h(a)= const$ then $z(\tau)= const$.

\medskip
\textbf{2.} $h(a)= \lambda a$, $\lambda \neq 0$

The second way to lift curve is lifting 
proportional to the length of the plane part, i.e. if
$h(a)= \lambda a$  then we have the following differential equitation on the `lifting' function $z(\tau)$
\[
\left( 1-\lambda^2\right) z_{\tau}^2 = \lambda^2 \left( x_{\tau}^2+y_{\tau}^2\right), 
\]
solving which given $ 1-\lambda^2>0$, we get 
\[
z(\tau)= \pm \frac{   \lambda  }{\sqrt{1-\lambda^2}}\, l(\tau) + C ,
\] 
where $ l(\tau) $ is length of plane projection of curve and C is a constant.

Here if $\lambda=\pm 1$ then $x(t)=y(t)=const$ and we have a vertical line.

\medskip
\textbf{3.} $h(a)= \lambda a^2$, $\lambda\neq 0$ 

In this case we have the following differential equation on the `lifting' function $z(\tau)$
\[
\left( 1- 4\lambda z\right) z_{\tau}^2= 4\lambda z\left( x_{\tau}^2+y_{\tau}^2\right) ,
\] 
solving which under assumption  
$
0<\lambda z<\frac{1}{4}, 
$
we get the following relation between the `lifting' function $z(\tau)$ and the  length of the plane  curve 
\[
\sqrt{4\lambda z (1-4\lambda z)} - \arccos(\sqrt{4\lambda z}) =\pm 4\lambda l(\tau).
\]

\begin{figure}[h]
	\centering
	\begin{subfigure}[b]{0.36\textwidth}
		\includegraphics[width=\textwidth]{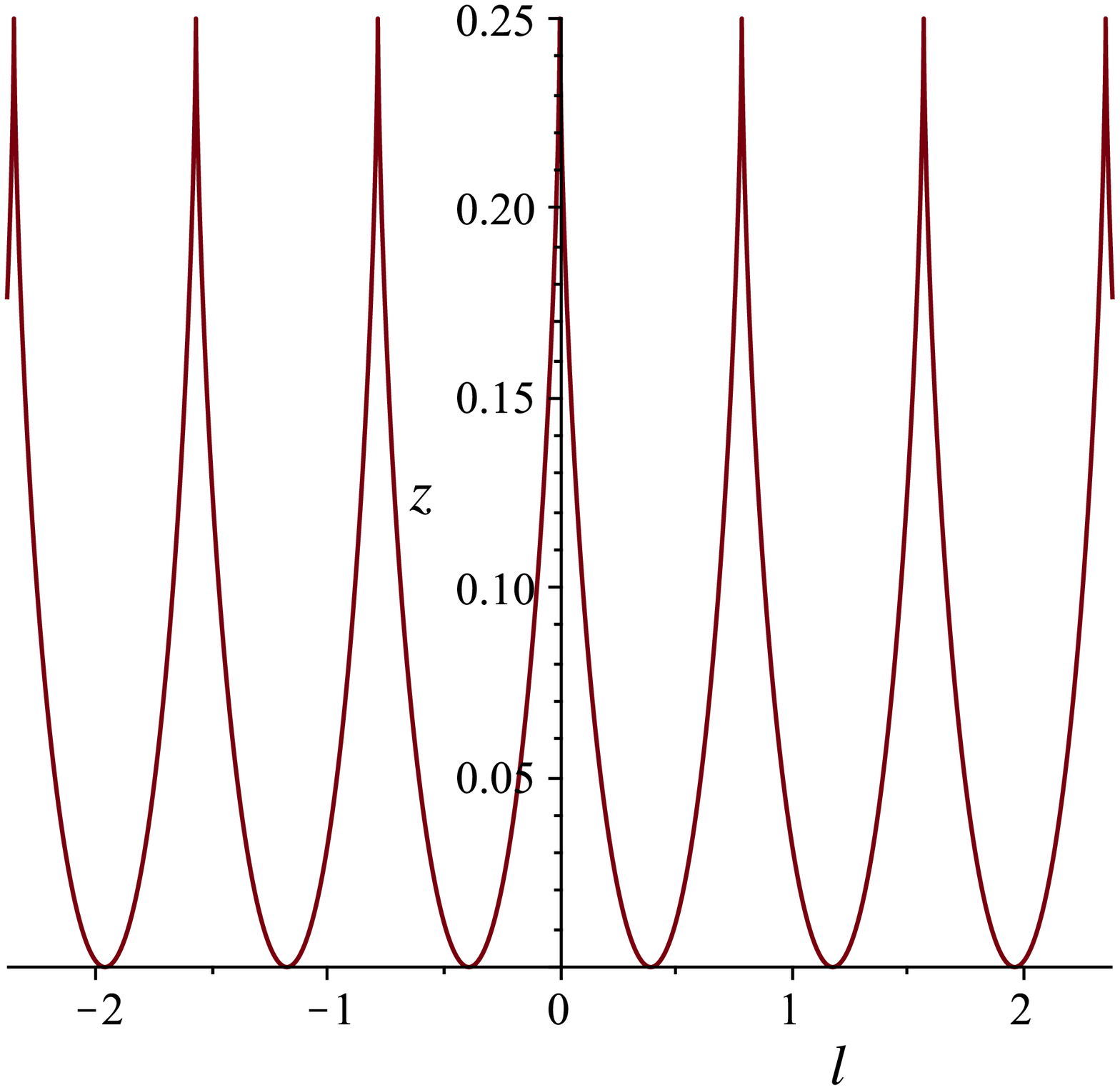}
		\caption{}\label{fig:pic3-1}
	\end{subfigure}  \qquad
	\begin{subfigure}[b]{0.40\textwidth}
		\centering
		\includegraphics[width=\textwidth]{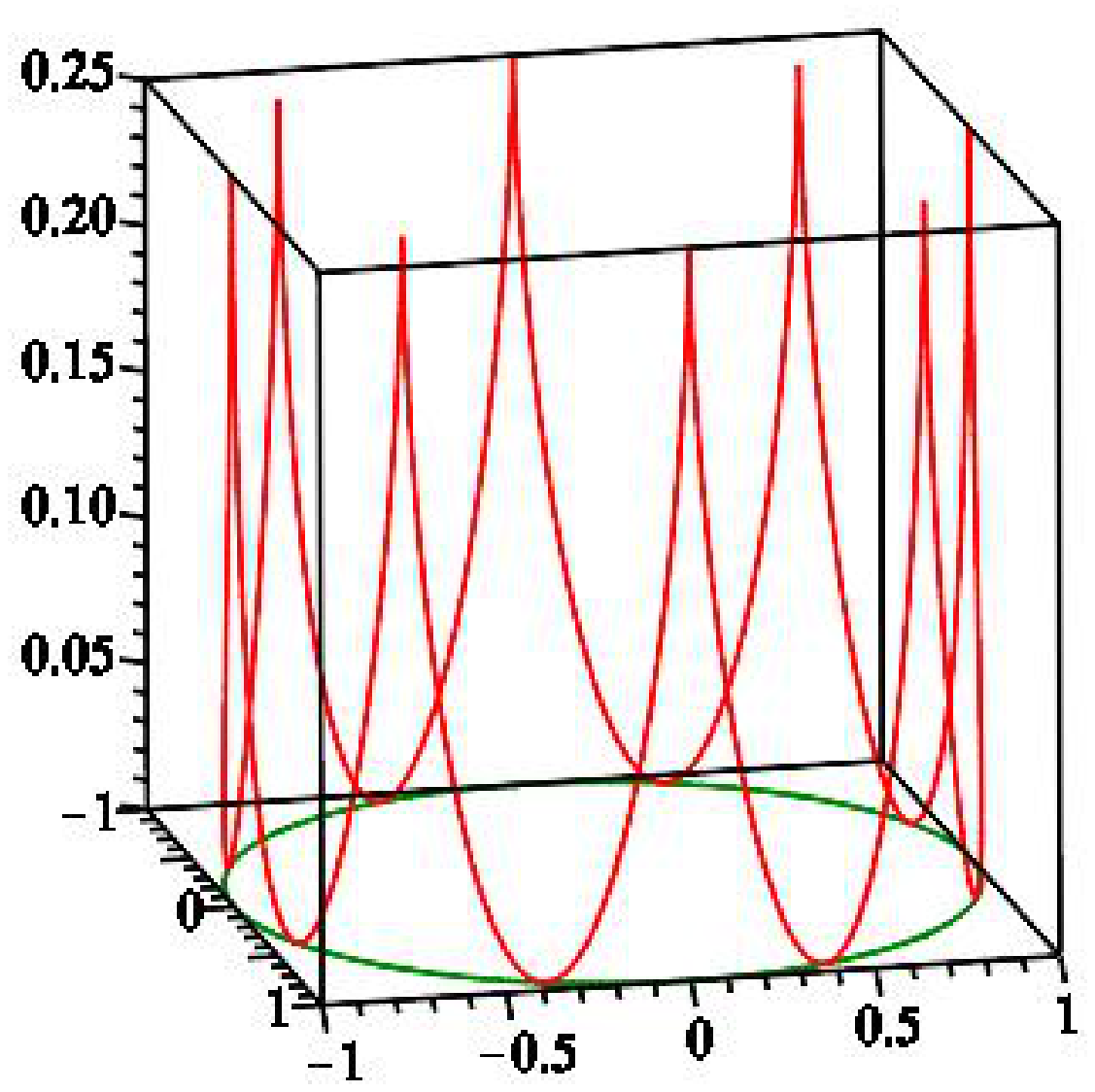}
		\caption{}\label{fig:pic3-2}
	\end{subfigure}
\caption{}
\end{figure}

The latter can be rewritten in the parametric form
\[
l=\pm \frac{\sin t \cos t - t}{4\lambda}, \quad
z=\frac{\cos^2 t}{4\lambda} ,
\]
which is useful to demonstrate relationship between $l(\tau)$ and $z(\tau)$ on a graph (see Figure \ref{fig:pic3-1}). 

Consider an example of lifting of a unit circle with the `lifting'
function we found (Figure \ref{fig:pic3-2}).

\medskip
\textbf{4.} $h(a)= \lambda_1a^{\lambda_2} $, $\lambda_2\neq 0,1,2$  

In this case the differential equation on the `lifting' function $z(\tau)$ is
\[ 
 z_t^2\left( \left(\frac{z}{\lambda_1}\right)^{\frac{2}{\lambda_2}} -\lambda_2^2 z^2\right)  =\lambda_2^2 z^2 \left( x_{\tau}^2+y_{\tau}^2\right) .
\]

General solution of this equation is expressed in terms of hypergeometric functions. For example, for the case $\lambda_1=1$, $\lambda_2=\frac{11}{3}$ we get

\[
z^{\frac{3}{11}}\, _2F_1\left(-\frac{1}{2},\frac{3}{16};\frac{19}{16};
\frac{121}{9}z^{\frac{16}{11}}\right) = \pm l(\tau).
\]

\medskip
\textbf{5.} $h(a)= \lambda_1e^{\lambda_2a} $  

In this case the differential equation on the `lifting' function $z(\tau)$ has the form
\[
z_t^2\left( 1-\lambda_2^2 z^2\right) = \lambda_2^2 z^2 \left( x_{\tau}^2+y_{\tau}^2\right),
\]
solving which under assumption $1-\lambda_2^2z^2>0 $, we get relation between the `lifting' function $z(\tau)$ and the length of the plane curve
\[
\sqrt{1-\lambda_2^2z^2}-\frac{1}{2}\ln\frac{1+\sqrt{1-\lambda_2^2z^2}}{1-\sqrt{1-\lambda_2^2z^2}} =\pm \lambda_2 l(\tau)
\]
or in the parametric form
\[
l = \pm \frac{1}{\lambda_2} \left( \sin t - \frac{1}{2}\ln\frac{1+\sin t}{1-\sin t}   \right), \quad
z=\dfrac{1}{\lambda_2}\cos t, \quad \cos t \neq 0.
\]

\begin{figure}[h]
	\centering
	\begin{subfigure}[b]{0.38\textwidth}
		\includegraphics[width=\textwidth]{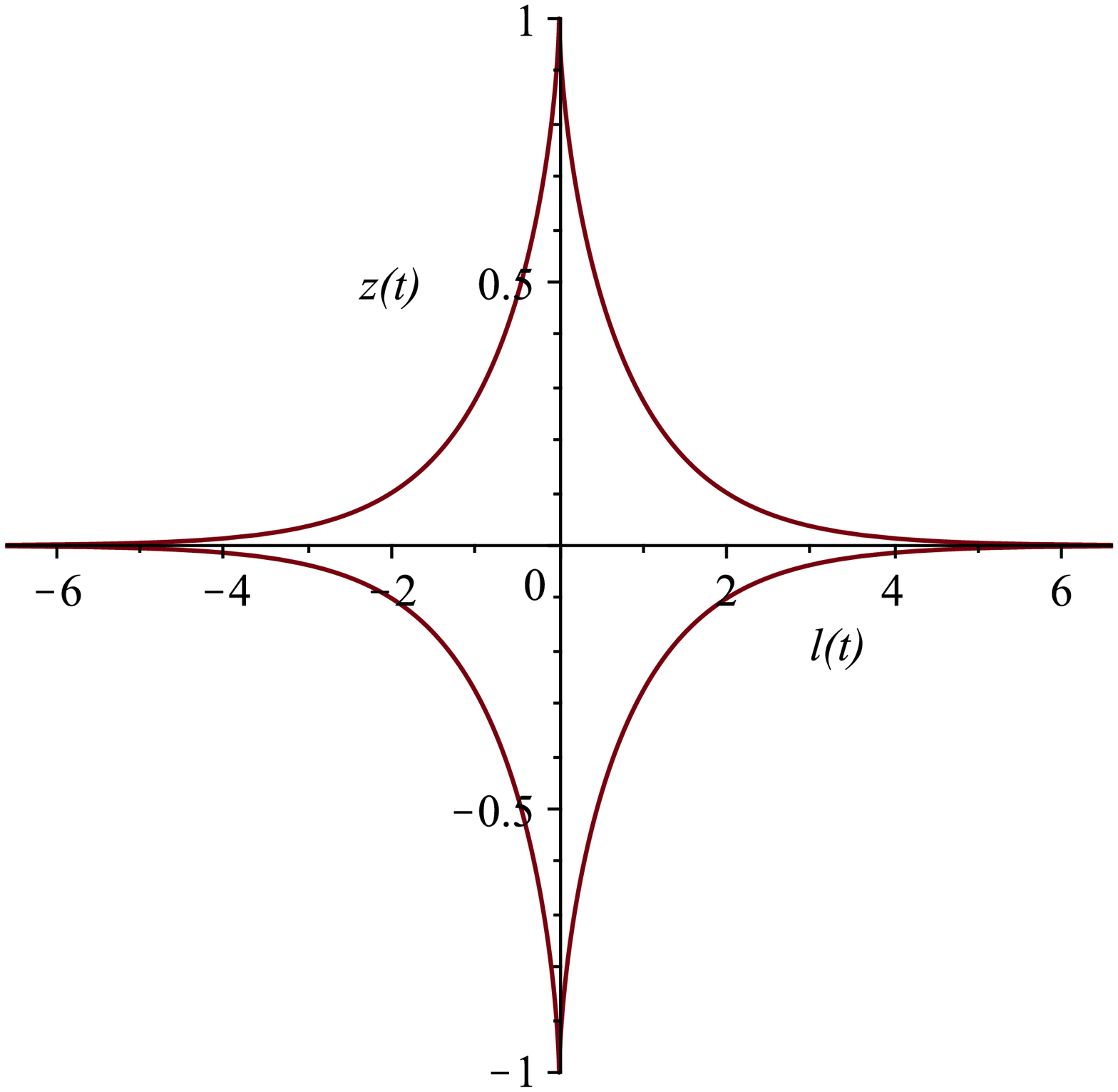}
		\caption{}\label{fig:pic5-1}
	\end{subfigure}  \qquad
	\begin{subfigure}[b]{0.40\textwidth}
		\centering
		\includegraphics[width=\textwidth]{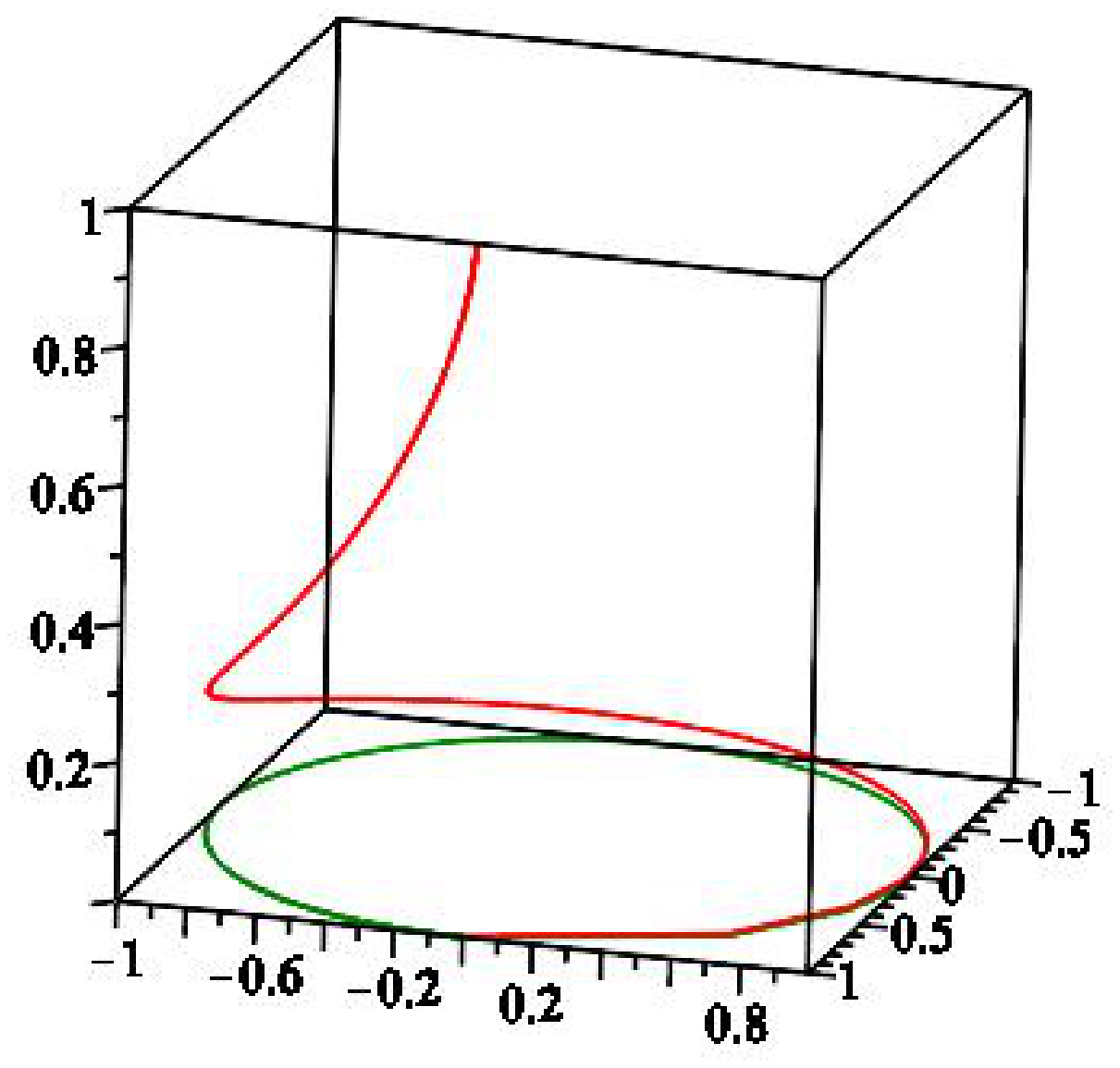}
		\caption{}\label{fig:pic5-2}
	\end{subfigure}
	\caption{}
\end{figure}

The relation between the plane curve
length $l$ and the `lifting' function $z$
is shown on Figure \ref{fig:pic5-1}. On Figure \ref{fig:pic5-2}
lifting of a unit circle is demonstrated, here we use positive values of $z$ and $l$. In fact, the space curve starts from the height of one unit above the circle and never intersects with it.

\medskip
\textbf{6.} $h(a)= \ln{a} $ 

In this case  the differential equation on the `lifting' function $z(\tau)$ has the form
\[
z_t^2\left( e^{2 z}-1\right)=\left( x_{\tau}^2+y_{\tau}^2\right),
\]
solving which under assumption $e^{2z}-1>0 $, we get relation between  functions $l(\tau)$ and $z(\tau)$
\[
\sqrt{ e^{2z}-1}-\arctan{\sqrt{ e^{2z}-1}} = \pm  l(\tau) .
\]
The parametric form of this solution is
\[
l = \pm ( \tan t - t ), \quad
z = -\ln|\cos t|, \quad
\tan t >0.
\]

\begin{figure}[h]
	\centering
	\begin{subfigure}[b]{0.36\textwidth}
		\includegraphics[width=\textwidth]{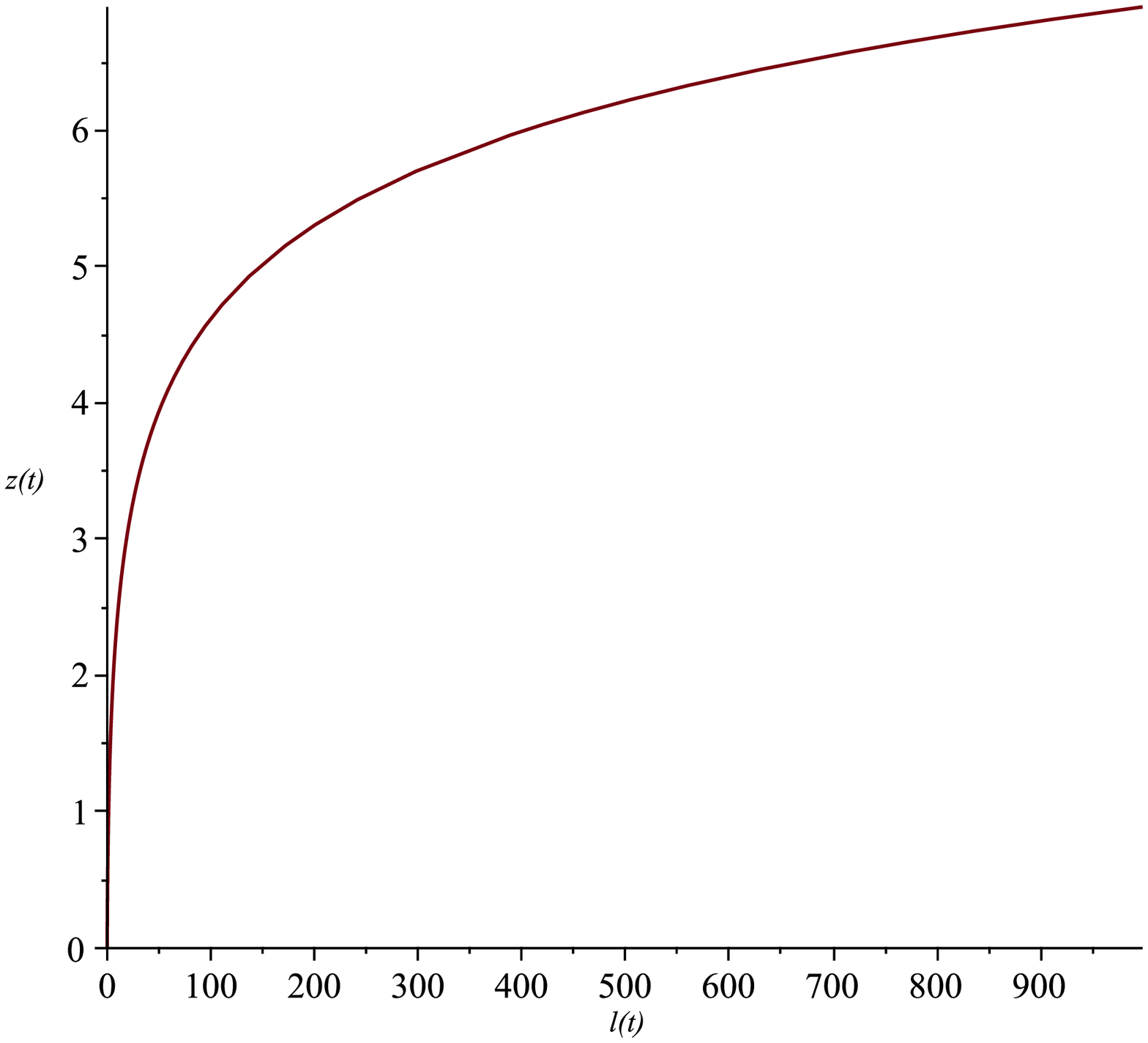}
		\caption{}\label{fig:pic6-1}
	\end{subfigure}  \qquad
	\begin{subfigure}[b]{0.40\textwidth}
		\centering
		\includegraphics[width=\textwidth]{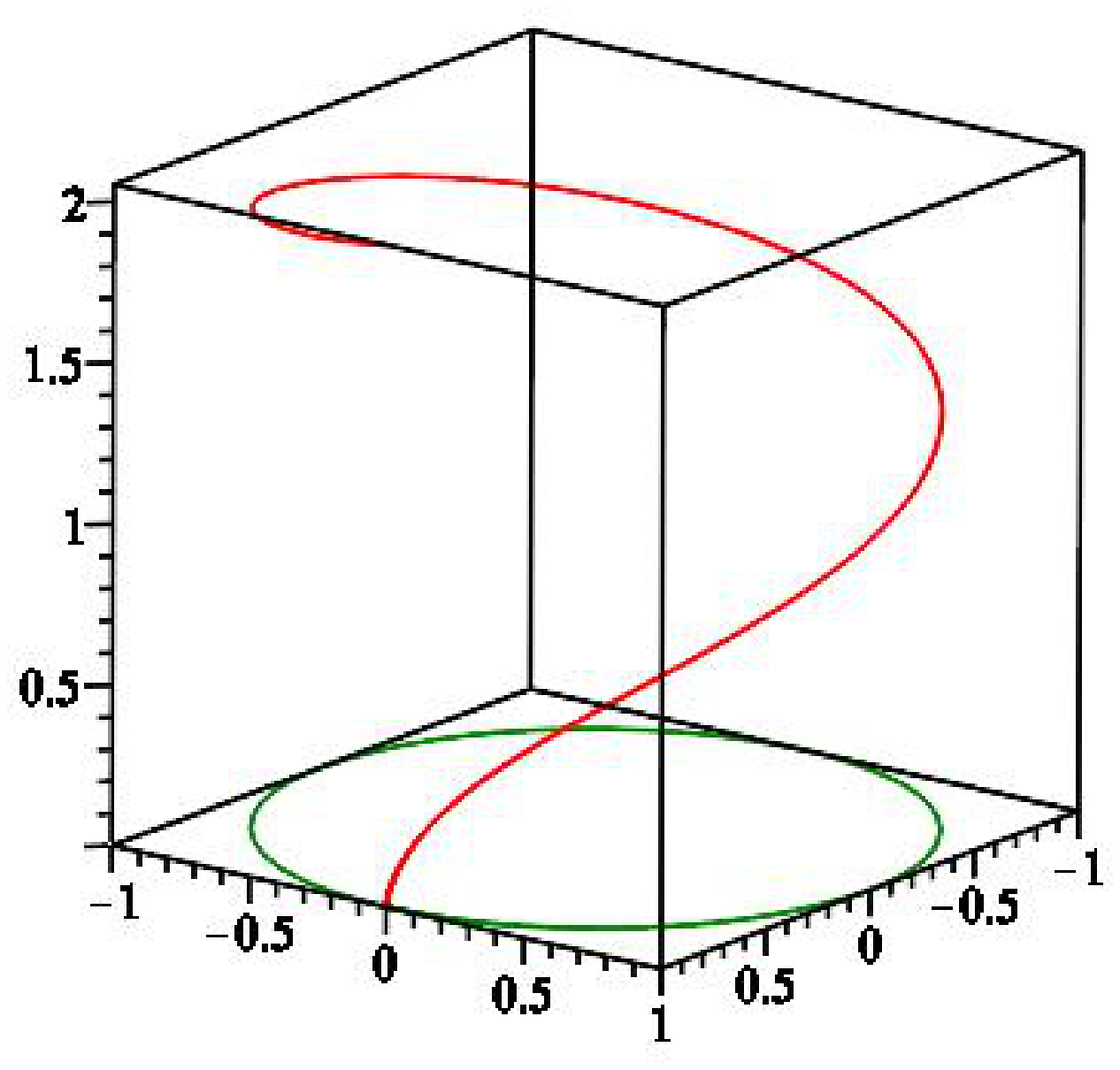}
		\caption{}\label{fig:pic6-2}
	\end{subfigure}
	\caption{}
\end{figure}

The relation between the plane curve
length $l$ and the `lifting' function $z$
is shown on Figure \ref{fig:pic6-1},
but only positive values of $l$
are plotted. On Figure \ref{fig:pic6-2}
the corresponding lifting of a unit circle is demonstrated.

\section{Thermodynamic states with a one-dimensional symmetry algebra} \label{sec:therm}

In this section we consider the thermodynamic states, or the Lagrangian surfaces $L$, admitting a one-dimensional symmetry algebra.
The cases, when thermodynamic states admit a two-dimensional symmetry algebra, can be studied in the similar manner.
  
First of all, consider the case, when $h(a)$ is an arbitrary function. 

Let the thermodynamic state admit a one-dimensional symmetry algebra. Denote by
\[
Z=\gamma_1\dir{ \press}+ \gamma_2\dir{ \entr}+ \gamma_3\temp\,\dir{ \temp} + \gamma_4( \press\,\dir{ \press} + \dens\,\dir{ \dens} -\entr\,\dir{ \entr})
\]
a basis vector of this algebra, then the Lagrangian surface can be found from the following PDE system on the internal energy (see \cite{DLTwisla} for details) 
\begin{equation*}\label{qwe1}
\left\{
\begin{aligned}
&\gamma_4\dens\,{\energy}_{\dens\dens} + ( \gamma_2-\gamma_4\entr ){\energy}_{\dens \entr}  + \gamma_4 {\energy}_{\dens}  - \frac{\gamma_1}{\dens^2} =0, \\
&(\gamma_2-\gamma_4\entr){\energy}_{\entr \entr} + \gamma_4\dens\,{\energy}_{\dens \entr} -\gamma_3\,{\energy}_{ \entr} =0.
\end{aligned} \right. 
\end{equation*}

It is easy to check that the Mayer bracket \cite{Kruglikov2002} of these two equations vanishes, and therefore the system is formally integrable and compatible.

Solving this system for the general case,  we find expressions for the pressure and the temperature 
\begin{equation*}
\press =C_1\dens - \frac{\gamma_1 }{\gamma_4}  ,\quad
\temp =C_2(\gamma_2-\gamma_4\entr)^{-\frac{\gamma_3}{\gamma_4}}   ,
\end{equation*}
where $C_1, C_2$ are constants.

Moreover, the negative definiteness of the quadratic differential form $\kappa$ on the Lagrangian surface leads to the relations
\[
\frac{\gamma_3}{\gamma_2-\gamma_4\entr} >0, \quad
C_1 >0
\]
for all $\entr \in(-\infty,\entr_0]$.

\begin{theorem}
	The thermodynamic states admitting a one-dimensional symmetry algebra  have the form 
	\begin{equation*}
	\press =C_1\dens - \frac{\gamma_1 }{\gamma_4}  ,\quad
	\temp =C_2(\gamma_2-\gamma_4\entr)^{-\frac{\gamma_3}{\gamma_4}} ,
	\end{equation*}
	where the constants defining the symmetry algebra satisfy inequalities 
\[
\entr_0<\frac{\gamma_2}{\gamma_4}, \quad
C_1>0, \quad   \frac{\gamma_1 }{\gamma_4}<0, 
\]
	and besides they must meet one of the following conditions:
	\begin{enumerate}
		\item if $\frac{\gamma_3}{\gamma_4}$ is irrational, then  $\gamma_3>0$, $\gamma_4>0$, $C_2>0$;
		\item if $\frac{\gamma_3}{\gamma_4}$ is rational, then $\frac{\gamma_3}{\gamma_4}>0$ (i.e. $\frac{\gamma_3}{\gamma_4}=\frac{m}{k}$ ) and
		\begin{enumerate}
			\item if $k$ is even, then $\gamma_4>0$, $C_2>0$;
			\item if $k$ is odd and $m$ is even, then $C_2>0$;
			\item if $k$ is odd and $m$ is odd, then $C_2\gamma_4>0$.
		\end{enumerate}
	\end{enumerate}
\end{theorem}

Below we consider the special cases.

\medskip
\textbf{1, 2.} $h(a)=const$, $h(a)=\lambda a$

The pure thermodynamic part of the system symmetry algebra coincides with the thermodynamic part of the 2d Euler case. So, the classification of the thermodynamic states for these two cases can be found in \cite{DLTwisla}.

\medskip
\textbf{3.} $h(a)= \lambda a^2$, $\lambda\neq 0$

Let a basis vector of a one-dimensional symmetry algebra be
\[
\gamma_1\dir{ \press} +\gamma_2\dir{ \entr}  + \gamma_3\temp\,\dir{ \temp} +\gamma_4 \dens\,\dir{ \dens}  + \gamma_5(\press\,\dir{ \press} - \entr\,\dir{ \entr}),
\]
then in the general case expressions for the pressure and temperature have the form
\begin{equation*}
\press =C_1\dens^{\frac{\gamma_5}{\gamma_4}} - \frac{\gamma_1 }{\gamma_5}  ,\quad
\temp =C_2(\gamma_2-\gamma_5\entr)^{-\frac{\gamma_3}{\gamma_5}}   ,
\end{equation*}
where $C_1, C_2$ are constants.
The admissibility conditions (the negative definiteness  of the differential form $\kappa$) lead to the relations 
\[
\frac{\gamma_3}{\gamma_2-\gamma_5\entr} >0, \quad
\frac{\gamma_5C_1}{\gamma_4} >0
\]
for all $\entr \in(-\infty,\entr_0]$.

\medskip
\textbf{4.} $h(a)= \lambda_1 a^{\lambda_2}$, $\lambda_1,\lambda_2\neq 0$ 

Let a basis vector of a one-dimensional symmetry algebra be
\[
\gamma_1\dir{ \press} +\gamma_2\dir{ \entr}  + \gamma_3\temp\,\dir{ \temp} +
\gamma_4 (2\lambda_2 \dens\,\dir{ \dens} -(\lambda_2-2) \entr\,\dir{ \entr})  + 
\gamma_5(\press\,\dir{ \press} + \dens\,\dir{ \dens} - \entr\,\dir{ \entr}),
\]
then in the general case expressions for the pressure and temperature have the form
\begin{equation*}
\press =C_1\dens^{\frac{\gamma_5}{2\lambda_2\gamma_4+\gamma_5}}     - \frac{\gamma_1 }{\gamma_5}  ,\quad
\temp =C_2 \left(\left((\lambda_2-2)\gamma_4+\gamma_5 \right) \entr - \gamma_2  \right) ^{-\frac{\gamma_3}{\left( \lambda_2-2\right)\gamma_4+\gamma_5 }}   ,
\end{equation*}
where $C_1, C_2$ are constants. The admissibility conditions (the negative definiteness  of the differential form $\kappa$) lead to the relations 
\[
\frac{\gamma_3}{\left((\lambda_2-2)\gamma_4+\gamma_5 \right) \entr - \gamma_2 } <0, \quad
\frac{\gamma_5C_1}{2\lambda_2\gamma_4+\gamma_5} >0
\]
for all $\entr \in(-\infty,\entr_0]$.

\medskip
\textbf{5.} $h(a)=\lambda_1e^{\lambda_2}$

The pure thermodynamic part of the system symmetry algebra coincides 
with the symmetry Lie algebra of the Euler system of differential equations on a two dimensional unit sphere. So, the classification of thermodynamic states can be found in \cite{DLTwisla}.

\medskip
\textbf{6.} $h(a)= \ln a$ 

Let a basis vector of a one-dimensional symmetry algebra be 
\[
\gamma_1\dir{ \press} +\gamma_2\dir{ \entr}  + \gamma_3\temp\,\dir{ \temp} +\gamma_4 \dens\,\dir{ \dens}  + \gamma_5(\press\,\dir{ \press} + \dens\,\dir{ \dens}),
\]
then in the general case expressions for the pressure and temperature have the form
\begin{equation*}
\press =C_1\dens - \frac{\gamma_1 }{\gamma_5}  ,\quad
\temp =C_2(\gamma_2+\gamma_4\entr)^{\frac{\gamma_3}{\gamma_4}}   ,
\end{equation*}
where $C_1, C_2$ are constants. The admissibility conditions (the negative definiteness  of the differential form $\kappa$) lead to the relations 
\[
\frac{\gamma_3}{\gamma_2+\gamma_4\entr} >0, \quad
C_1 >0
\]
for all $\entr \in(-\infty,\entr_0]$.


\section{ Differential invariants} \label{sec:invs}

As in \cite{DLTwisla}, we consider two group actions
on the Euler system $\systemEk{}$.
Namely, the prolonged actions of the groups generated
by actions of the Lie algebras $\LieAlgebra{g_{m}}$ and 
$\LieAlgebra{g_{sym}}$.

Recall that 
a function $J$ on the manifold $\systemEk{k}$
is a \textit{kinematic differential invariant of order} $\leq k$ if 
\begin{enumerate}
	\item $J$ is a rational function along fibers of the projection $\pi_{k,0}:\systemEk{k}\rightarrow \systemEk{0}$,
	\item $J$ is invariant with respect to the prolonged action of the Lie algebra $\LieAlgebra{g_{m}}$, i.e., for all $X\in \LieAlgebra{g_{m}}$,
	\begin{equation} \label{dfinv}
	X^{(k)}(J)=0,
	\end{equation}
\end{enumerate}
where $\systemEk{k}$ is the prolongation of the system $\systemEk{}$ to $k$-jets,
and $X^{(k)}$ is the $k$-th prolongation of a vector field $X\in \LieAlgebra{g_{m}}$.

Note that fibers of the projection $\systemEk{k}\rightarrow \systemEk{0}$ are irreducible algebraic manifolds.

A kinematic invariant is \textit{an Euler invariant} if condition \eqref{dfinv} holds for all $X \in \LieAlgebra{g_{sym}}$.

We say that a point $x_k\in \systemEk{k}$ and the corresponding orbit $\mathcal{O}(x_k)$ ($\LieAlgebra{g_{m}}$- or $\LieAlgebra{g_{sym}}$-orbit) are \textit{regular}, if there are exactly $m=\mathrm{codim}\, \mathcal{O}(x_k) $ independent  invariants (kinematic or Euler) in a neighborhood of this orbit. 
Otherwise, the point and the corresponding orbit are \textit{singular}.

The Euler system together with the symmetry algebras
$ \LieAlgebra{g_{m}}$ or $\LieAlgebra{g_{sym}}$ satisfies the conditions
of Lie-Tresse theorem (see \cite{KL}), and therefore
the kinematic 
and Euler differential invariants separate
regular $ \LieAlgebra{g_{m}}$ and $\LieAlgebra{g_{sym}}$
orbits on the Euler system $\systemEk{}$ correspondingly. 

By a $\LieAlgebra{g_{m}}$ or $\LieAlgebra{g_{sym}}$-invariant derivation we mean a total derivation
\[
A\totalDiff{t}+B\totalDiff{a}
\]
that commutes with prolonged action of algebra $\LieAlgebra{g_{m}}$  or $\LieAlgebra{g_{sym}}$.
Here $A$, $B$ are rational functions on the prolonged system $\systemEk{k}$ for some $k\geq 0$.

\subsection{Kinematic invariants}

\begin{theorem} 
	\begin{enumerate}
		\item The kinematic invariants field is generated
		by first-order basis differential invariants and by basis invariant derivations. This field separates regular orbits.
		\item For the general cases of $h(a)$, as well as for  $h(a)=\lambda_1a^{\lambda_2}$, $h(a)=\lambda_1e^{\lambda_2a}$ and $h(a)=\ln a$, the basis differential invariants are
		\[
		a,\quad u,\quad \dens,\quad\entr,\quad u_a,\quad\dens_a,\quad \entr_t, \quad\entr_a,
		\]
		and the basis invariant derivations are
		\[
		\totalDiff{t} ,\quad \totalDiff{a} . 
		\]
		\item For the cases $h(a)=const$, $h(a)=\lambda a$ and $h(a)=\lambda a^2$, the basis differential invariants are
		\[
		\dens,\quad\entr,\quad u_a, \quad\dens_a, \quad \entr_a, \quad \entr_t+u \entr_a,
		\]
		and basis invariant derivations are
		\[
		 \totalDiff{t} + u \totalDiff{a} ,\quad  \totalDiff{a}.
		\]	
		\item The number of independent invariants of pure order $k$	is equal to $4$ for $k\geq 1$.
	\end{enumerate}
\end{theorem}

\subsection{Euler invariants}

First of all, consider the case, when $h(a)$ is an arbitrary function.

Let the thermodynamic state admit a one-dimensional symmetry algebra generated by the vector field
\begin{equation*}
A =  \xi_1 X_2 + \xi_2 X_3 + \xi_3 X_4 + \xi_4 X_5.
\end{equation*}

The action of the thermodynamic vector field $A$ on the field of kinematic invariants is given by the derivation
\[
(\xi_2-\xi_4\entr)\dir{\entr} + \xi_4(\dens\dir{\dens} +\dens_a \dir{\dens_a} -\entr_t\dir{\entr_t} - \entr_a\dir{\entr_a}  ), 
\]
finding first integrals of this vector field we get basis Euler invariants of the first order. 
\begin{theorem} 
	The field of Euler differential invariants for thermodynamic states
	admitting a one-dimensional symmetry algebra is generated by
	the differential invariants
\[
a, \quad \left( \entr - \dfrac{\gamma_2}{\gamma_4}\right) \dens , \quad
u, \quad u_a, \quad \frac{\dens_a}{\dens} , \quad \entr_t\dens,  \quad \entr_a \dens
\]
of the first order and by the invariant derivations 
\[
 \totalDiff{t} ,\quad   \totalDiff{a} .
\]
This field separates the regular orbits.
\end{theorem}

Below the special cases for the function $h(a)$ are considered.

\medskip
\textbf{1.} $h(a)=const$

If the thermodynamic state admits a one-dimensional symmetry algebra generated by the vector field
\begin{equation*}
\xi_1 X_2 + \xi_2 X_3 + \xi_3 X_4 + \xi_4 X_5 +\xi_5 X_8 + \xi_6 X_9,
\end{equation*}
then the field of Euler differential invariants is generated by
the differential invariants
\[
\left( \entr - \dfrac{\xi_2}{\xi_4+\xi_5-\xi_6}\right) \frac{\dens_a}{\dens\entr_a}, \quad
\frac{u_a\dens^{\frac{\xi_6}{\xi_4}+1}}{\dens_a}, \quad
\dens_a\dens^{\frac{\xi_5}{\xi_4}-1}, \quad
\frac{u_a\entr_a\dens^4}{\dens_a^3}, \quad
\frac{(\entr_t+u \entr_a)\dens^3}{\dens_a^2}
\]
of the first order and by the invariant derivations 
\[
\dens^{\frac{\xi_5+\xi_6}{\xi_4}}\left( \totalDiff{t} + u \totalDiff{a}\right)  ,\quad 
 \dens^{\frac{\xi_5}{\xi_4}}\totalDiff{a} .
\]

\medskip
\textbf{2.}  $h(a)=\lambda a$, $\lambda\neq 0$

If the thermodynamic state admits a one-dimensional symmetry algebra generated by the vector field
\begin{equation*}
 \xi_1 X_2 + \xi_2 X_3 + \xi_3 X_4 + \xi_4 X_5 +\xi_5 X_8 + \xi_6 X_9,
\end{equation*}
then the field of Euler differential invariants is generated by
the differential invariants
\[
\left( \entr - \dfrac{\xi_2}{\xi_4+\xi_5}\right) \frac{\dens_a}{\dens\entr_a}, \quad
u_a\dens^{\frac{\lambda\mathrm{g}\xi_5}{\lambda\mathrm{g}(\xi_4-2\xi_5)-2\xi_6}}, \quad
\dens_au_a^{-2}\dens^{\frac{\xi_6}{\lambda\mathrm{g}(\xi_4-2\xi_5)-2\xi_6}-1}, \quad
\frac{u_a\entr_a\dens^4}{\dens_a^3}, \quad
\frac{(\entr_t+u \entr_a)\dens^3}{\dens_a^2}
\]
of the first order and by the invariant derivations 
\[
 \dens^{\frac{\lambda\mathrm{g}\xi_5}{\lambda\mathrm{g}(\xi_4-2\xi_5)-2\xi_6}}\left( \totalDiff{t} + u \totalDiff{a}\right)  ,\quad 
 \dens^{\frac{2\lambda\mathrm{g}\xi_5+\xi_6}{\lambda\mathrm{g}(\xi_4-2\xi_5)-2\xi_6}} \totalDiff{a} .
\]

\medskip
\textbf{3.}  $h(a)=\lambda a^2$, $\lambda\neq 0$

If the thermodynamic state admits a one-dimensional symmetry algebra generated by the vector field
\begin{equation*}
 \xi_1 X_2 + \xi_2 X_3 + \xi_3 X_4 + \xi_4 X_5 +\xi_5 X_6,
\end{equation*}
then the field of Euler differential invariants is generated by
the differential invariants
\[
\left( \entr - \dfrac{\xi_2}{\xi_4}\right) \frac{\dens_a}{\dens\entr_a}, \quad u_a, \quad
\dens_a\dens^{\frac{\xi_5}{\xi_4-2\xi_5}-1}, \quad
\frac{\entr_a\dens^4}{\dens_a^3}, \quad
\frac{(\entr_t+u \entr_a)\dens^3}{\dens_a^2}
\]
of the first order and by the invariant derivations 
\[
 \totalDiff{t} + u \totalDiff{a}  ,\quad 
 \dens^{\frac{\xi_5}{\xi_4-2\xi_5}}\totalDiff{a} .
\]

\medskip
\textbf{4.}  $h(a)= \lambda_1a^{\lambda_2}$, $\lambda\neq 0, 1, 2$

If the thermodynamic state admits a one-dimensional symmetry algebra generated by the vector field
\begin{equation*}
 \xi_1 X_2 + \xi_2 X_3 + \xi_3 X_4 + \xi_4 X_5 +\xi_5 X_6,
\end{equation*}
then the field of Euler differential invariants is generated by
the differential invariants
\[
\dens u^2 a^{\frac{\xi_4(\xi_2-2)}{2\xi_5}},\quad
\left(\entr - \frac{\xi_2}{\xi_4+\xi_5} \right)a u\dens, \quad ua^{-\frac{\xi_2}{2}}, \quad \frac{u_a a}{u}, \quad 
\frac{\dens_a a}{\dens}, \quad
\entr_ta^2\dens, \quad 
\entr_a a^2  u \dens 
\]
of the first order and by the invariant derivations 
\[
a^{1-\frac{\xi_2}{2}} \totalDiff{t} ,\quad  a \totalDiff{a} .
\]

\medskip
\textbf{5.}  $h(a)= \lambda_1e^{\lambda_2a}$

If the thermodynamic state admits a one-dimensional symmetry algebra generated by the vector field
\begin{equation*}
 \xi_1 X_2 + \xi_2 X_3 + \xi_3 X_4 + \xi_4 X_5 +\xi_5 X_6,
\end{equation*}
then the field of Euler differential invariants is generated by
the differential invariants
\[
\dens u e^{\frac{\xi_2\xi_4a}{2\xi_5}},\quad
\left(\entr - \frac{\xi_2}{\xi_4} \right) u\dens, \quad ue^{-\frac{\xi_2a}{2}}, \quad \frac{u_a }{u}, \quad 
\frac{\dens_a }{\dens}, \quad
\entr_t\dens, \quad 
\entr_a u \dens 
\]
of the first order and by the invariant derivations 
\[
 e^{-\frac{\xi_2a}{2}} \totalDiff{t} ,\quad   \totalDiff{a} .
\]

\medskip
\textbf{6.}  $h(a)=\ln a$

If the thermodynamic state admits a one-dimensional symmetry algebra generated by the vector field
\begin{equation*}
 \xi_1 X_2 + \xi_2 X_3 + \xi_3 X_4 + \xi_4 X_5 +\xi_5 X_6,
\end{equation*}
then the field of Euler differential invariants is generated by
the differential invariants
\[
\dens a^{-\frac{\xi_4}{\xi_5}},\quad
\left(\entr - \frac{\xi_2}{\xi_4+\xi_5} \right) a\dens, \quad u, \quad u_a a, \quad 
\frac{\dens_a a}{\dens}, \quad
\entr_ta^2\dens, \quad 
\entr_a a^2  \dens 
\]
of the first order and by the invariant derivations 
\[
a\totalDiff{t} ,\quad a\totalDiff{a} .
\]

\newpage
\section*{Appendix}

In the table below we summarize the connection
between the function $h$ and
the symmetry Lie algebra of the system \eqref{eq:E},
see Section \ref{sec:symmetries} for details.

\vspace*{12pt}
\begingroup
\setlength{\tabcolsep}{2pt}
\renewcommand{\arraystretch}{2.9}
\begin{tabular}{||b{0.3\linewidth}|b{0.6\linewidth}||}
	\hline
	$h(a)$ is arbitrary
	&$\begin{aligned} 
	&X_1 = \dir{ t}, \qquad\\
	&X_2 = \dir{\press }, \qquad \\
	&X_3 = \dir{\entr}, \\
	&X_4 = \temp\,\dir{ \temp}, \\
	&X_5 =  \press\,\dir{ \press} + \dens\,\dir{ \dens} - \entr\,\dir{ \entr}
	\end{aligned} $ \\
	\hline
	$h(a)= const$ 
	&$\begin{aligned} 
	&X_6 = \dir{ a}, \qquad\\
	&X_7 = t\,\dir{ a}+\dir{ u}, \qquad \\
	&X_8 = t\,\dir{ t}+a\,\dir{a}-\entr\,\dir{\entr}, \\
	&X_{9} = t\,\dir{ t} - u\,\dir{ u} -2 \press\,\dir{ \press} + \entr\,\dir{ \entr}
	\end{aligned} $ \\
	\hline
	$h(a)= \lambda a$, $\lambda\neq 0 $ 
	& $\begin{aligned} 
	&X_6 = \dir{ a}, \qquad\\
	&X_7 = t\,\dir{ a}+\dir{ u}, \qquad \\
	&X_8 = t\,\dir{ t}+2a\,\dir{a} +  u\,\dir{ u} -2 \dens\,\dir{ \dens} -\entr\,\dir{\entr}, \\
	&X_{9} = (\frac{t^2}{2}+\frac{a}{\lambda g})\,\dir{ a} +2(t+ \frac{u}{\lambda g})\,\dir{ u} -\frac{2 \dens}{\lambda g}\,\dir{ \dens} 
	\end{aligned} $ \\
	\hline
	$h(a)= \lambda a^2$, $\lambda\neq 0 $
	&$\begin{aligned} 
	&X_6 = a \dir{ a} +u\,\dir{ u}  -2 \dens\,\dir{ \dens} , \qquad\\
	&X_7 = \sin(\sqrt{2\lambda g}\,t)\,\dir{a} +  \sqrt{2\lambda g} \cos(\sqrt{2\lambda g}\,t)\,\dir{ u} , \qquad \\
	&X_8 =\cos(\sqrt{2\lambda g}\,t)\,\dir{a} -  \sqrt{2\lambda g} \sin(\sqrt{2\lambda g}\,t)\,\dir{ u}  \\
	\end{aligned} $\\
	\hline
	$h(a)= \lambda_1a^{\lambda_2}$, $\lambda_2\neq 0,1,2 $
	&$\begin{aligned} 
	&X_6 = t\,\dir{ t} -\frac{2a}{\lambda_2-2} \dir{ a} - \frac{\lambda_2 u }{\lambda_2-2}\dir{ u} +
	\frac{2\lambda_2 \dens }{\lambda_2-2}\dir{ \dens} - \entr\,\dir{ \entr}   \qquad
	\end{aligned}$\\
	\hline
	$h(a)= \lambda_1e^{\lambda_2a}$, $\lambda_2\neq 0 $
	&$\begin{aligned} 
	&X_6 = t\,\dir{ t}-\frac{2	}{\lambda_2}\,\dir{a} -  u\,\dir{ u} - \press\,\dir{ \press} +\dens\,\dir{\dens}
	\end{aligned}$\\
	\hline
	$h(a)= \ln{a}$ 
	&$\begin{aligned} 
	&X_6  = t\,\dir{ t}+ a  \,\dir{a}-\entr\,\dir{\entr}
	\end{aligned}$\\
	\hline
\end{tabular}
\vspace{12pt}
\endgroup

\textbf{Acknowledgments.} The research was partially supported by RFBR Grant No 18-29-10013.


\newpage
\begin{thebibliography}{99}

\bibitem{Anderson2016} Anderson, Ian M. and Torre, Charles G., {\it The Differential Geometry Package} (2016). Downloads. Paper 4. 
http://digitalcommons.usu.edu/dg\_downloads/4

\bibitem{Batchelor2000} Batchelor G. K.
{\it An introduction to fluid dynamics.} Cambridge university press, 2000.

\bibitem{DLTwisla}
Duyunova A., Lychagin V., Tychkov S. {\it Differential invariants for  flows of fluids and gases.} ArXiv:2004.01567 [math-ph].


\bibitem{KL} Kruglikov~B., Lychagin~V., {\it Global Lie-Tresse theorem}. Selecta Math. 2016, 22, 1357-1411.

\bibitem{Kruglikov2002} 
B. Kruglikov, V. Lychagin, {\it Mayer brackets and solvability	of PDEs--I}. Differential Geometry and its Applications, Elsevier BV, 17, 251-272 (2002).


\end{thebibliography}
\end{document}